\documentclass[%
 reprint,
 amsmath,amssymb,
 aps,
prb,
 longbibliography
]{revtex4-2}

\usepackage[dvipdfmx]{graphicx}
\usepackage{dcolumn}
\usepackage{bm}
\usepackage{color}
\usepackage[usenames]{xcolor}
\usepackage{siunitx}
\usepackage{makecell}

\begin{document}

\title{Electronic and magnetic properties of iridium ilmenites $A$IrO$_3$ ($A=$ Mg, Zn, and Mn)}

\author{Seong-Hoon Jang}
\author{Yukitoshi Motome}

\affiliation{
	Department of Applied Physics, University of Tokyo, Tokyo 113-8656, Japan
}

\date{\today}

\begin{abstract}
We theoretically investigate the electronic band structures and magnetic properties of ilmenites with edge-sharing IrO$_6$ honeycomb layers, $A$IrO$_3$ with $A=$ Mg, Zn, and Mn, in comparison with a collinear antiferromagnet MnTiO$_3$. 
The compounds with $A=$ Mg and Zn were recently reported in Y.~Haraguchi {\it et al.}, Phys. Rev. Materials {\bf 2}, 054411 (2018), while MnIrO$_3$ has not been synthesized yet but the honeycomb stacking structure was elaborated in a superlattice with MnTiO$_3$ in K.~Miura {\it et al.}, Commun. Mater. {\bf 1}, 55 (2020). 
We find that, in contrast to MnTiO$_3$, where an energy gap opens in the Ti $3d$ bands by antiferromagnetic ordering of the high-spin $S=5/2$ moments, MgIrO$_3$ and ZnIrO$_3$ have a gap in the Ir $5d$ bands under the influence of both spin-orbit coupling and electron correlation. 
Their electronic structures are similar to those in the spin-orbit coupled Mott insulators with the $j_{\rm eff}=1/2$ pseudospin degree of freedom, as found in monoclinic $A_2$IrO$_3$ with $A=$ Na and Li which have been studied as candidates for the Kitaev spin liquid. 
Indeed, we find that the effective exchange interactions between the $j_{\rm eff}=1/2$ pseudospins are dominated by the Kitaev-type bond-dependent interaction and the symmetric off-diagonal interactions. 
On the other hand, for MnIrO$_3$, we show that the local lattice structure is largely deformed, and both Mn $3d$ and Ir $5d$ bands appear near the Fermi level in a complicated manner, which makes the electronic and magnetic properties qualitatively different from MgIrO$_3$ and ZnIrO$_3$.
Our results indicate that the IrO$_6$ honeycomb network in the ilmenites $A$IrO$_3$ with $A=$ Mg and Zn would offer a good platform for exotic magnetism by the spin-orbital entangled moments like the Kitaev spin liquid.
\end{abstract}


\maketitle

\section{Introduction}
\label{sec:introduction}

\begin{figure}[th!]
\includegraphics[width=0.9\columnwidth]{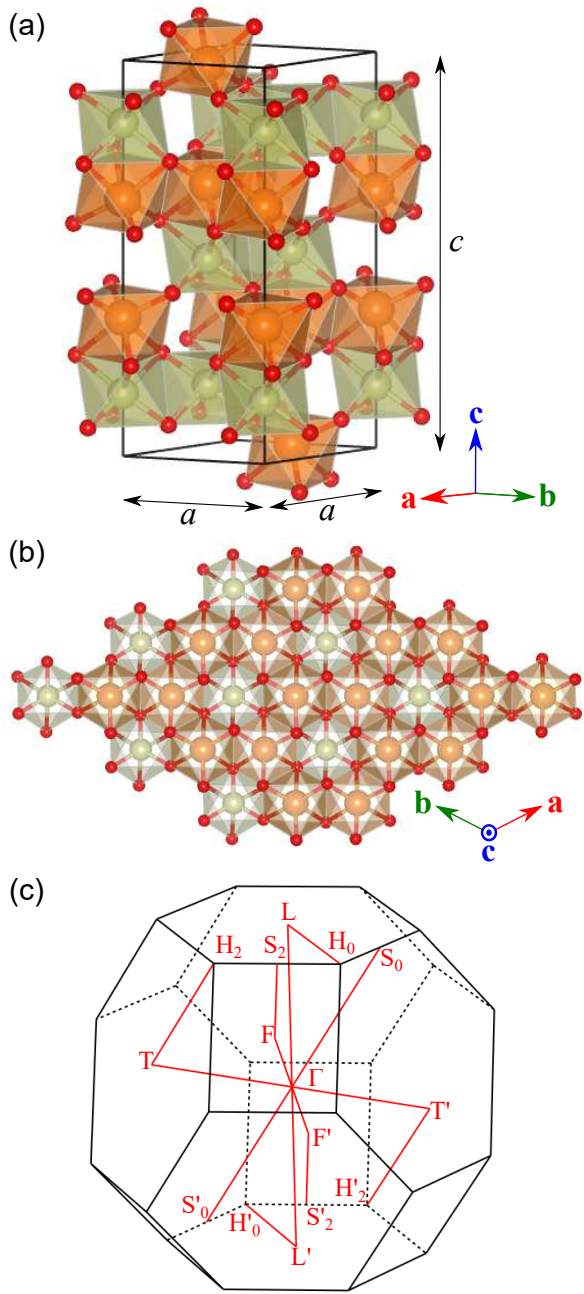} 
\caption{\label{fig:ilmenite}
(a) and (b) Lattice structure of ilmenite $AB$O$_3$. 
The orange and yellow octahedra denote $A$O$_6$ and $B$O$_6$, each of which form a two-dimensional honeycomb network with edge sharing. 
The adjacent $A$O$_6$ and $B$O$_6$ honeycomb layers are stacked with face and corner sharing. 
(a) Bird's-eye view and (b) projection from the $c$ axis. 
The black lines in (a) denote the conventional unit cell with the lattice constants, $a$ and $c$.
(c) The first Brillouin zone.
The red lines denote the symmetric lines used in the plots of the band structures in Sec.~\ref{sec:result} and Appendix~\ref{appA}.
}
\end{figure} 

Ilmenite, whose chemical formula is given by $AB$O$_3$, crystalizes in a trigonal structure with the space group $R\bar{3}$ similar to corundum $A_2$O$_3$. 
Both ilmenite and corundum share the layered structure with a honeycomb network of edge-sharing octahedra, but a difference lies in the stacking manner; corundum is composed of a stacking of isostructural $A$O$_6$ honeycomb layers, but ilmenite is made of an alternative stacking of $A$O$_6$ and $B$O$_6$ honeycomb layers, as shown in Fig.~\ref{fig:ilmenite}. 
While ilmenite is originally the name for a titanium-iron oxide mineral FeTiO$_3$, its relatives, such as NiTiO$_3$, CoTiO$_3$, and MnTiO$_3$, have been studied for a long time as a good playground for two-dimensional magnetism~\cite{doi:10.1143/JPSJ.13.1110, doi:10.1063/1.1729357, Newnham:a04103, Shirane1959, PhysRev.164.765, AKIMITSU197087, doi:10.1143/JPSJ.42.462, WATANABE1980549, YAMAGUCHI1986865, PhysRevB.101.195122,PhysRevLett.57.483}. 
Mixed compounds like (Ni,Mn)TiO$_3$ were also investigated as they exhibit interesting spin glass behavior~\cite{PhysRevLett.57.483, PhysRevLett.59.2364, doi:10.1143/JPSJ.58.1416, doi:10.1143/JPSJ.62.2575, doi:10.1143/JPSJ.63.3145}. 
Later, the titanium antiferromagnets have also attracted the interest from their multiferroics behavior~\cite{PhysRevLett.108.057203,PhysRevB.93.104404} and magnetochiral dichroism~\cite{PhysRevLett.124.217402}.

Recently, a new series of ilmenite with $B$=Ir has been synthesized as MgIrO$_3$, ZnIrO$_3$, and CdIrO$_3$~\cite{PhysRevMaterials.2.054411, PhysRevMaterials.4.044401}. 
These compounds are of particular interest from a different perspective than $A$TiO$_3$: they have a honeycomb network of edge-sharing IrO$_6$ octahedra similar to monoclinic $A_2$IrO$_3$ with $A=$ Na and Li which have been intensively studied as candidates for realizing a quantum spin liquid in the honeycomb Kitaev model~\cite{PhysRevLett.105.027204, SI2010, SI2012, PhysRevB.88.035107, Katukuri_2014, PhysRevLett.113.107201, WI2016, Freund2016, WI2017, TA2019, doi:10.7566/JPSJ.89.012002}. 
In $A_2$IrO$_3$, the $5d$ levels in Ir$^{4+}$ ions are split by the strong spin-orbit coupling (SOC) into quartet and doublet, and the half-filled doublet is further split by the Coulomb interaction to realize the so-called spin-orbit Mott insulator~\cite{PhysRevLett.101.076402}. 
Then, the low-energy physics is described by the pseudospin degree of freedom for the Kramers doublet with the effective magnetic moment of $j_{\rm eff}=1/2$~\cite{PhysRevLett.109.266406, PhysRevB.88.085125}. 
Owing to the edge-sharing geometry, the dominant interaction between neighboring $j_{\rm eff}=1/2$ moments can be highly anisotropic, which gives a realization of the bond-dependent Ising interaction in the Kitaev model~\cite{HwanChun2015, PhysRevB.99.081101, PhysRevB.88.035107, Katukuri_2014, PhysRevLett.113.107201, WI2016}. 
Since the iridium ilmenites have a similar honeycomb network, they potentially serve as another candidates for the Kitaev spin liquid.
Powder samples of these compounds, however, were shown to exhibit magnetic phase transitions at $31.8$~K for MgIrO$_3$, $46.6$~K for ZnIrO$_3$~\cite{PhysRevMaterials.2.054411}, and $90.9$~K in CdIrO$_3$~\cite{PhysRevMaterials.4.044401}, which are higher than $\sim 15$~K for $A$IrO$_3$~\cite{SI2010, SI2012, PhysRevB.85.180403, Williams2016, PhysRevB.83.220403}. 
The susceptibility measurements for the $A=$ Mg and Zn indicate that they have in-plane magnetic anisotropy, while Na$_2$IrO$_3$ and Li$_2$IrO$_3$ show the out-of-plane and in-plane anisotropy, respectively~\cite{SI2010, Freund2016}. 
The estimates of the magnetic moments are consistent with the $j_{\rm eff}=1/2$ picture, except for CdIrO$_3$~\cite{PhysRevMaterials.4.044401}. 
Despite these interesting aspects, the electronic and magnetic properties of the iridium ilmenites have not been theoretically studied thus far. 

In this paper, we investigate the electronic band structures of MgIrO$_3$ and ZnIrO$_3$ by using the first-principles calculations with the fully-relativistic local density approximation including effective onsite Coulomb interactions, called the LDA+SOC+$U$ method. 
For comparison, we study the well-known antiferromagnetic insulator MnTiO$_3$ and a fictitious crystal MnIrO$_3$ whose local stacking structure was recently elaborated in a superlattice with MnTiO$_3$~\cite{Miura2020}. 
We find that MgIrO$_3$ and ZnIrO$_3$ have similar band structures near the Fermi level to the Kitaev candidates Na$_2$IrO$_3$ and Li$_2$IrO$_3$; the SOC and Coulomb interactions act cooperatively to realize the spin-orbit coupled Mott insulating state whose low-energy physics is well described by the pseudospin with effective magnetic moment $j_{\rm eff}=1/2$.
This is in contrast to the antiferromagnetic insulating state in MnTiO$_3$, where $(3d)^5$ electrons form the high-spin $S=5/2$ state by the Hund's-rule coupling and the energy gap is opened by the exchange splitting in the antiferromagnetic state. 
In MgIrO$_3$ and ZnIrO$_3$, we show that the antiferromagnetic solution has a lower energy than the paramagnetic and ferromagnetic ones, but the antiferromagnetic moment is very small $\sim 0.1$~$\mu_{\rm B}$, implying that the $j_{\rm eff}=1/2$ moments suffer from frustration.
Furthermore, by constructing a multiorbital Hubbard model from the maximally-localized Wannier functions (MLWFs)~\cite{Wannier1937, MO2014} and performing the perturbation expansion from the atomic limit, we show that the exchange interactions between the $j_{\rm eff}=1/2$ pseudospins are described by the dominant Kitaev-type bond-dependent one and the subdominant symmetric off-diagonal ones. 
The results indicate that the edge-sharing honeycomb network of IrO$_3$ octahedra in MgIrO$_3$ and ZnIrO$_3$ would offer a good playground for spin-orbital entangled magnetism toward the Kitaev spin liquid.
On the other hand, we find that the optimized lattice structure of MnIrO$_3$ is largely deformed from those for MnTiO$_3$, and the band structure near the Fermi level is complicated including both Mn $3d$ and Ir $5d$ contributions.

The structure of this paper is as follows. 
In Sec.~\ref{sec:method}, we describe the details of the LDA+SOC+$U$ calculations and the method to estimate the effective exchange coupling constants. 
In Sec.~\ref{sec:result}, we present our results for MnTiO$_3$ (Sec.~\ref{subsec:MnTiO3}), MgIrO$_3$ and ZnIrO$_3$ (Sec.~\ref{subsec:MgTiO3 and ZnIrO3}), and MnIrO$_3$ (Sec.~\ref{subsec:MnIrO3}). 
In Sec.~\ref{subsec:MgTiO3 and ZnIrO3}, we discuss the electronic band structure in Sec.~\ref{subsubsec:electronic structure}, the transfer integrals in Sec.~\ref{subsubsec:transfer}, and the effective magnetic interactions in Sec.~\ref{subsubsec:effective interaction}. 
The results for ZnIrO$_3$ are qualitatively similar to those for MgIrO$_3$, and detailed in Appendix~\ref{appA}. 
Section~\ref{sec:summary} is devoted to the summary.

\section{Method}
\label{sec:method}

The {\it ab initio} calculations are performed by using \texttt{Quantum ESPRESSO}~\cite{GI2017}. 
We adopt the fully-relativistic and non-relativistic projector-augmented-wave-method Perdew-Zunger type pseudopotentials for the $A$ and $B$-site ions and the O ligands, respectively~\cite{PE1981, BL1994, DalCorso2014}. 
While we employ the experimental structural data for MnTiO$_3$~\cite{Liferovich2005} and for MgIrO$_3$ and ZnIrO$_3$~\cite{PhysRevMaterials.2.054411}, we perform structural optimization for the fictitious compound MnIrO$_3$ starting from the experimental structure for MnTiO$_3$ with replacement of Ti by Ir; we relax not only the atomic positions within the primitive unit cell but also the lattice translation vectors.
In the optimization, we set the minimum ionic displacement to 0.001~\si{\angstrom} in the Broyden-Fletcher-Goldfarb-Shanno iteration scheme~\cite{BR1970}. 
Afterwards, we symmetrize the optimal structure within the trigonal space group $R\bar{3}$, where the residual stress is less than $30$~kbar. 
In all the calculations, we take the primitive unit cell, and $4\times4\times4$ and $8\times8\times8$ Monkhorst-Pack $\mathbf{k}$-grids for self-consistent field and non self-consistent field calculations, respectively~\cite{MO1976}.
We set the convergence threshold for the self-consistent field calculations to $10^{-10}$~Ry.
The kinetic energy cutoff is set to $200$~Ry for all the cases.
In the LDA+SOC+$U$ calculations, we include the Hubbard correction to the Mn $3d$ and Ir $5d$ orbitals with the Coulomb repulsion $U=U_{\rm Mn}$ and $U_{\rm Ir}$, respectively, together with the Hund's-rule coupling $J_{\rm H}$, by assuming $J_{\rm H}/U = 0.1$ in the rotationally invariant scheme~\cite{PhysRevB.52.R5467}. 

We construct the MLWFs of Ir $5d$ $t_{2g}$ and O $2p$ orbitals for MgIrO$_3$ and of Zn $3d$, Ir $5d$ $t_{2g}$, and O $2p$ orbitals for ZnIrO$_3$ for the obtained electronic band structures by using \texttt{WANNIER90}~\cite{MO2014}.
Note that, in most of the previous studies for other Kitaev candidate materials, the MLWF analyses were performed only for the $5d$ or $4d$ $t_{2g}$ orbitals~\cite{PhysRevB.88.035107, PhysRevLett.113.107201, KI2016, WI2016,Sugita2020}. 
In the present study, however, we include O $2p$ for both MgIrO$_3$ and ZnIrO$_3$ and also Zn $3d$, since we find that they overlap with Ir $5d$ $t_{2g}$ (see Sec.~\ref{subsec:MgTiO3 and ZnIrO3} and Appendix~\ref{appA}). 
From the results, we calculate the projected density of states (DOS) for each MLWF orbital. 
For the Ir $5d$ $t_{2g}$ orbitals, we also compute the DOS projected onto the spin-orbital coupled bases labelled by the effective angular momentum $j_{\rm eff}$. 
Then, we estimate the effective transfer integrals between the Ir $t_{2g}$ orbitals by using the MLWFs for the $U_{\rm Ir}=0$ case, including both $d$-$d$ direct and indirect contributions via the O $2p$ orbitals. 
The latter indirect contributions are obtained by the perturbation theory in terms of the $d$-$p$ transfer integrals with the use of the energy differences between the $d$ and $p$ levels, $\Delta_{d-p}$, as the intermediate-state energies. 
We also perform the calculations by taking into account the Coulomb interaction in the $2p$ orbitals, $U_p$, by replacing $\Delta_{d-p}$ with $\Delta_{d-p}+U_p$, considering less screening in the present MLWF analyses including the O $2p$ orbitals (see Sec.~\ref{subsubsec:effective interaction}). 
Finally, we construct the multiorbital Hubbard models for the Ir $t_{2g}$ orbitals and perform the perturbation expansion from the atomic limit to derive the effective Hamiltonian for the $j_{\rm eff}=1/2$ pseudospins of the Ir ions. 
The scheme is basically common to Refs.~\onlinecite{JA2019, JA2020}. 

\section{Result}
\label{sec:result}

\subsection{MnTiO$_3$}
\label{subsec:MnTiO3}

Before going into the iridium ilmenites, we start with the well-known MnTiO$_3$ as a reference.
This compound is an antiferromagnetic insulator with a collinear N\'{e}el order along the $c$ axis, which we call the $c$-AFM state hereafter~\cite{Shirane1959}. 
The energy gap is estimated as $\simeq 3.18$~eV~\cite{Enhessari2012}, and the magnetic moment is estimated as $\simeq 4.55$~$\mu_{\rm B}$~\cite{Shirane1959}, which is consistent with the high-spin state of Mn$^{3+}$ ions with $S=5/2$ under the strong Hund's-rule coupling. 
The electronic band structure was studied by the {\it ab initio} calculations with the generalized gradient approximation, and the AFM nature was reproduced~\cite{Deng2012}.

\begin{figure}[t]
\includegraphics[width=0.85\columnwidth]{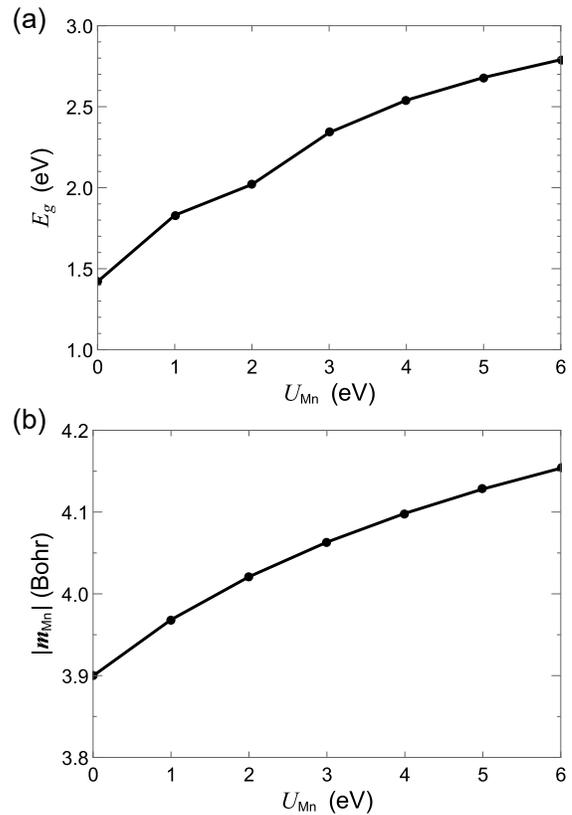} 
\caption{\label{fig:MnTiO3_Udep}
(a) The energy gap and (b) the antiferromagnetic moment at the Mn site in the $c$-AFM state in MnTiO$_3$ as functions of the Coulomb repulsion at the Mn site, $U_{\rm Mn}$, obtained by the LDA+SOC+$U$ calculations.}
\end{figure} 

We here perform the LDA+SOC+$U$ method while changing the Coulomb repulsion at the Mn site, $U_{\rm Mn}$. 
We find that the antiferromagnetic state has a lower energy compared to the paramagnetic and ferromagnetic solutions, while the energy does not depend so much on the direction of the antiferromagnetic moments (the energy difference between the states with in-plane and out-of-plane moments is smaller than $0.1$~meV for all $U_{\rm Mn}$). 
Hence, in the following calculations, we assume the $c$-AFM state which is observed experimentally~\cite{Shirane1959}. 
The $c$-AFM state is insulating even in the absence of $U_{\rm Mn}$; 
the energy gap $E_g$ and the magnitude of the antiferromagnetic moment $|\mathbf{m}_{\rm Mn}|$ are estimated as $\simeq 1.4$~eV and $\simeq 3.9$~$\mu_{\rm B}$, respectively, at $U_{\rm Mn}=0$. 
Both $E_g$ and $|\mathbf{m}_{\rm Mn}|$ increase with $U_{\rm Mn}$, as plotted in Fig.~\ref{fig:MnTiO3_Udep}. 
We note that both values of $E_g$ and $|\mathbf{m}_{\rm Mn}|$ are slightly smaller than the experimental estimates but approach them for large $U_{\rm Mn}$. 
The large value of $|\mathbf{m}_{\rm Mn}|$ indicates that the antiferromagnetic moment is composed of the high-spin $S=5/2$ state of the Mn ions under the strong Hund's-rule coupling, consistent with the experiment.

\begin{figure}[t]
\includegraphics[width=1.0\columnwidth]{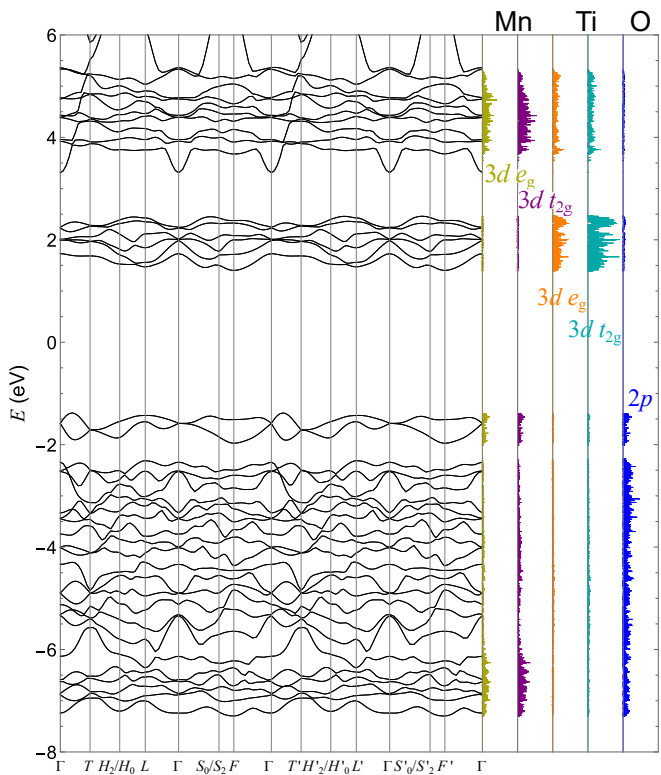} 
\caption{\label{fig:MnTiO3_band}
Electronic band structure of MnTiO$_3$ obtained by the LDA+SOC+$U$ calculations for the $c$-AFM state at $U_{\rm Mn}=6$~eV. 
The right panels display the projected DOS for the relevant orbitals in each ion. 
The Fermi level is set to zero. 
}
\end{figure} 

Figure~\ref{fig:MnTiO3_band} shows the electronic band structure of MnTiO$_3$ in the $c$-AFM state with $U_{\rm Mn}=6$~eV. 
The energy gap opens between the occupied states dominated by the Mn $3d$ and O $2p$ hybridized bands and the unoccupied states dominated by the Ti $3d$ bands. 
See the projected DOS in the right panels of Fig.~\ref{fig:MnTiO3_band}. 
The Mn $3d$ bands are largely split by the exchange energy from the $c$-AFM order.

\subsection{MgIrO$_3$ and ZnIrO$_3$}
\label{subsec:MgTiO3 and ZnIrO3}

\subsubsection{Electronic structure} 
\label{subsubsec:electronic structure}

Let us turn to the iridium ilmenites MgIrO$_3$ and ZnIrO$_3$. 
Since the two compounds have similar electronic band structures, we focus on MgIrO$_3$ in this section and present the results for ZnIrO$_3$ in Appendix~\ref{appA}. 

\begin{figure}[t]
\includegraphics[width=0.85\columnwidth]{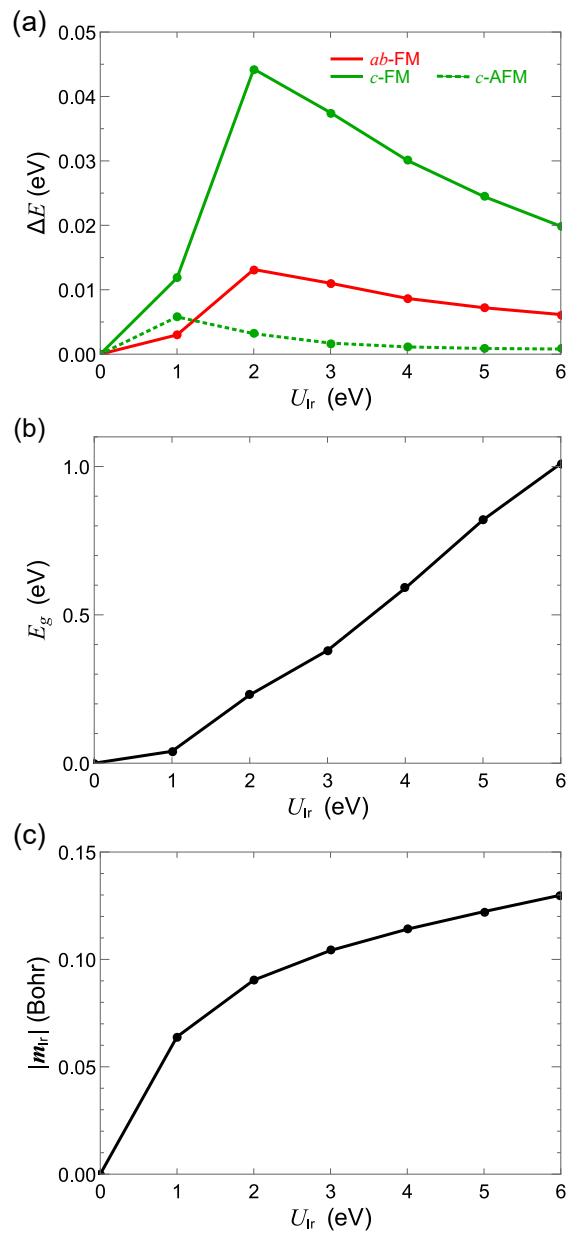} 
\caption{\label{fig:MgIrO3_Udep}
(a) Energy measured from the $ab$-AFM state in MgIrO$_3$ as a function of the Coulomb repulsion at the Ir site, $U_{\rm Ir}$, obtained by the LDA+SOC+$U$ calculations.
(b) The energy gap and (c) the antiferromagnetic moment at the Ir site in the $ab$-AFM state.
}
\end{figure} 

In MgIrO$_3$, the lowest-energy state among the different magnetic states which we calculated is the antiferromagnetic state whose moments lie in the $ab$ plane. 
In principle, the energy depends on the direction of the magnetic moments within the plane, but we do not find any significant energy change by rotating the direction (the energy difference between the states whose moments are parallel and perpendicular to one of the Ir-Ir bond directions is less than $0.2$~meV for all values of the Coulomb interaction at the Ir site, $U_{\rm Ir}$, calculated here). 
Hence, we measure the energy from the state with moments parallel to the bonds, which we call the $ab$-AFM state, and plot the result in Fig.~\ref{fig:MgIrO3_Udep}(a). 
We find that the $ab$-AFM state has the lowest energy in the whole range of $U_{\rm Ir}$, except for $U_{\rm Ir}=0$ where the system is a paramagnetic metal (see below). 
The result is consistent with the experiment where the magnetic susceptibility shows the easy-plane anisotropy~\cite{PhysRevMaterials.2.054411}. 
We note, however, that the energy difference between the $ab$-AFM and $c$-AFM state is not large and becomes smaller for larger $U_{\rm Ir}$.

\begin{figure}[t]
\includegraphics[width=1.0\columnwidth]{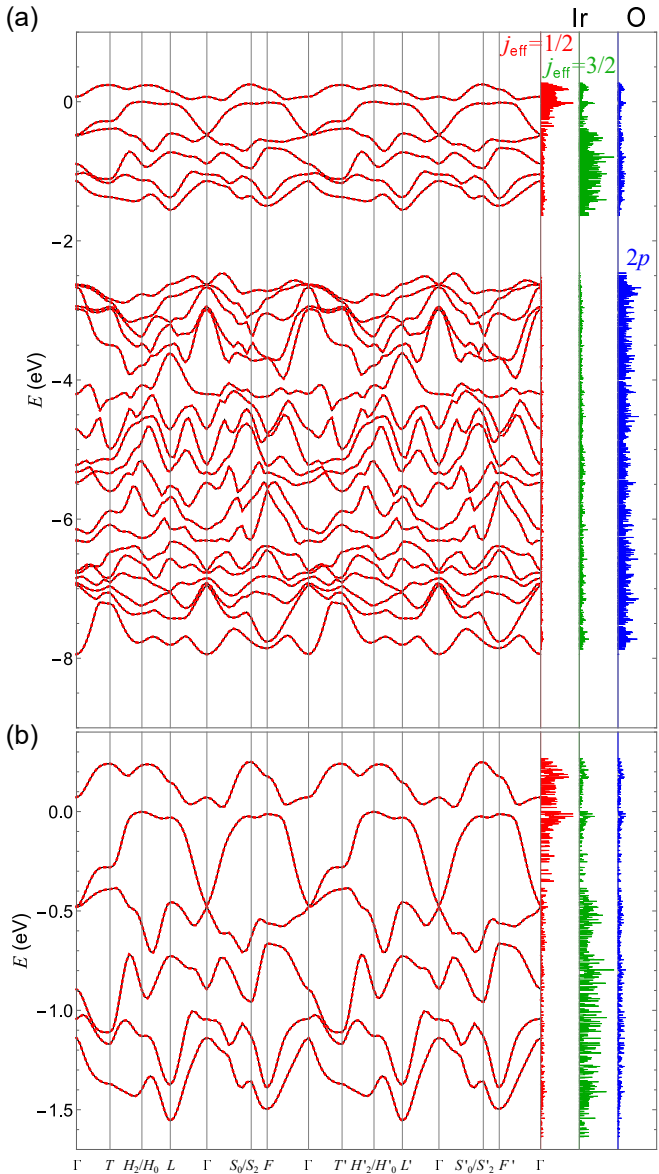}
\caption{\label{fig:MgIrO3_band_U=0}
The electronic band structure of MgIrO$_3$ obtained by the LDA calculations ($U_{\rm Ir}=0$) for the paramagnetic metallic state. 
The black curves denote the LDA results and the red dashed ones represent the band dispersions obtained by tight-binding parameters estimated by the MLWFs. 
The right panels display the projected DOS for each orbital. 
The Fermi level is set to zero. 
}
\end{figure} 

In Figs.~\ref{fig:MgIrO3_Udep}(b) and \ref{fig:MgIrO3_Udep}(c), we show the results of the energy gap $E_g$ and the magnitude of the magnetic moment of the Ir ion, $|\mathbf{m}_{\rm Ir}|$, as functions of $U_{\rm Ir}$. 
When $U_{\rm Ir}=0$, we obtain $E_g=0$ and $|\mathbf{m}_{\rm Ir}|=0$, indicating that the system is a paramagnetic metal. 
The band structure is shown in Fig.~\ref{fig:MgIrO3_band_U=0}. 
The relevant bands near the Fermi level are dominated by the Ir $5d$ states, which are composed of the lower-energy $j_{\rm eff}=3/2$ and higher-energy $j_{\rm eff}=1/2$ states split by the SOC, as shown in the projected DOS in the right panels of Fig.~\ref{fig:MgIrO3_band_U=0}. 
The Fermi level lies in the $j_{\rm eff}=1/2$ bands; the two bands in the $j_{\rm eff}=1/2$ manifold overlap slightly near the Fermi level, forming the metallic state, as shown in the enlarged plot in Fig.~\ref{fig:MgIrO3_band_U=0}(b). 

\begin{figure}[t]
\includegraphics[width=1.0\columnwidth]{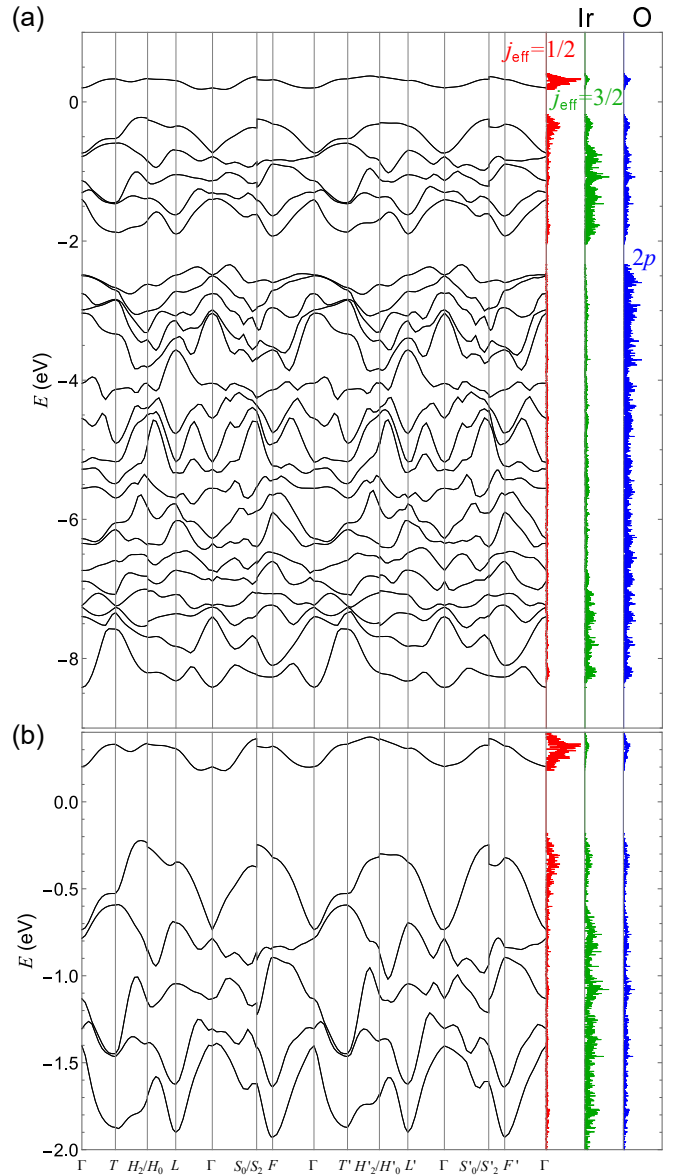} 
\caption{\label{fig:MgIrO3_band_U=3}
The electronic band structure of MgIrO$_3$ obtained by the LDA+SOC+$U$ calculations for the $ab$-AFM state with $U_{\rm Ir}=3$~eV. 
The notations are common to those in Fig.~\ref{fig:MgIrO3_band_U=0}.
}
\end{figure} 

When we switch on $U_{\rm Ir}$, the system turns into the $ab$-AFM insulating state, as shown in Figs.~\ref{fig:MgIrO3_Udep}(b) and \ref{fig:MgIrO3_Udep}(c). 
While $E_g$ increases almost linearly with $U_{\rm Ir}$, the gap value is relatively small compared to that for the $3d$ compound MnTiO$_3$ in Fig.~\ref{fig:MnTiO3_Udep}(a). 
In addition, $|\mathbf{m}_{\rm Ir}|$ grows slowly with $U_{\rm Ir}$ and has a small value of $|\mathbf{m}_{\rm Ir}| \simeq 0.1$~$\mu_{\rm B}$. 
The band structure of the $ab$-AFM insulating state is shown in Fig.~\ref{fig:MgIrO3_band_U=3} at $U_{\rm Ir}=3$~eV. 
In the $ab$-AFM state, the energy gap is opened by $U_{\rm Ir}$ between the two $j_{\rm eff}=1/2$ bands, while the $j_{\rm eff}=3/2$ bands slightly hybridize with them. 
This is a typical electronic band structure of the spin-orbit coupled Mott insulator, common to the Kitaev candidate materials like $A_2$IrO$_3$ ($A=$ Na and Li) and $\alpha$-RuCl$_3$~\cite{PhysRevLett.109.266406, PhysRevB.88.085125, PhysRevB.88.035107, PhysRevB.90.041112, JO2015, WI2017, C7CP07953B, Clancy2018}. 
The results suggest that the small magnetic moment in the $ab$-AFM state arises from the spin-orbit entangled moments described by the $j_{\rm eff}=1/2$ pseudospin degree of freedom. 

ZnIrO$_3$ shows similar behaviors; see Appendix~\ref{appA}. 
A difference from MgIrO$_3$ observed in our calculations is that the $ab$-AFM state has a slightly higher energy than the $c$-AFM state, which is not compatible with the experiment~\cite{PhysRevMaterials.2.054411}. 
Nonetheless, the electronic band structure indicates that this compound is also categorized into a spin-orbit coupled Mott insulator with the active $j_{\rm eff}=1/2$ pseudospins, similar to MgIrO$_3$. 

\subsubsection{Transfer integrals}
\label{subsubsec:transfer}

\begin{widetext}
\begin{table*}[th!]
\caption{\label{tab:tMgIrO3}
Transfer integrals between the Ir $t_{2g}$ orbitals on the nearest-neighbor $z$ bond for MgIrO$_3$. 
The values represent the effective transfer integrals from the orbital and spin in the column to those in the row, which are estimated from the MLWF analysis for the band structure at $U_{\rm Ir} = 0$ in Fig.~\ref{fig:MgIrO3_band_U=0}.
In each matrix element, we display three values by assuming $U_p=0.0$, $0.5$, and $1.0$~eV from top to bottom; see the text for details.
The unit is in meV. The upper-right half of the table is omitted as the matrix is Hermite conjugate.
}
\begin{ruledtabular}
\begin{tabular}{ccccccc}
&$yz\uparrow$&$yz\downarrow$&$zx\uparrow$&$zx\uparrow$&$xy\uparrow$&$xy\uparrow$\\
\hline
$yz\uparrow$&\makecell{101\\126\\138}\\
\hline
$yz\downarrow$&\makecell{0.00\\0.00\\0.00}&\makecell{101\\126\\138}\\
\hline
$zx\uparrow$&\makecell{1450-13.3{\rm i}\\904-12.0{\rm i}\\646-11.1{\rm i}}&\makecell{2.73{\rm i}\\2.13{\rm i}\\1.72{\rm i}}&\makecell{101\\126\\138}\\
\hline
$zx\downarrow$&\makecell{2.73{\rm i}\\2.13{\rm i}\\1.72{\rm i}}&\makecell{1450+13.3{\rm i}\\904+12.0{\rm i}\\646+11.1{\rm i}}&\makecell{0.00\\0.00\\0.00}&\makecell{101\\126\\138}\\
\hline
$xy\uparrow$&\makecell{28.0+11.4{\rm i}\\27.3+7.18{\rm i}\\24.5+5.09{\rm i}}&\makecell{-85.2{\rm i}\\-65.6{\rm i}\\-54.5{\rm i}}&\makecell{28.0+11.4{\rm i}\\27.3+7.18{\rm i}\\24.5+5.09{\rm i}}&\makecell{85.2{\rm i}\\65.6{\rm i}\\54.5{\rm i}}&\makecell{-423\\-438\\-450}\\
\hline
$xy\downarrow$&\makecell{-85.2{\rm i}\\-65.6{\rm i}\\-54.5{\rm i}}&\makecell{28.0-11.4{\rm i}\\27.3-7.18{\rm i}\\24.5-5.09{\rm i}}&\makecell{85.2{\rm i}\\65.6{\rm i}\\54.5{\rm i}}&\makecell{28.0-11.4{\rm i}\\27.3-7.18{\rm i}\\24.5-5.09{\rm i}}&\makecell{0.00\\0.00\\0.00}&\makecell{-423\\-438\\-450}\\
\end{tabular}
\end{ruledtabular}
\end{table*}
\end{widetext}

To examine whether the iridium ilmenites have dominant Kitaev-type bond-dependent interactions between the $j_{\rm eff}=1/2$ pseudospins, we first perform the MLWF analysis for the case of MgIrO$_3$ by using the band structure at $U_{\rm Ir} = 0$ in Fig.~\ref{fig:MgIrO3_band_U=0}. 
We find that the tight-binding model obtained from the MLWF analysis well reproduce the {\it ab initio} results, as shown in Fig.~\ref{fig:MgIrO3_band_U=0}. 
Then, following the procedures in Sec.~\ref{sec:method}, we estimate the effective transfer integrals between the Ir $t_{2g}$ orbitals, including both direct and indirect contributions. 
We present the results for the nearest-neighbor $z$ bond, where the effective Kitaev interaction takes the form of $S_i^z S_j^z$ [see Eq.~\eqref{eq:Heff} below], in Table~\ref{tab:tMgIrO3}.
Here, we show the estimates obtained by assuming $U_p=0.0$, $0.5$, and $1.0$~eV.
The values on the $x$ and $y$ bonds are obtained by cyclic permutations of $\{ xyz \}$. 

As shown in Table~\ref{tab:tMgIrO3}, we find that the most dominant transfer integral is the one between the $yz$ and $zx$ orbitals, which plays an important role in generating the Kitaev-type interaction~\cite{JA2009}.  
We note that the value at $U_p=0$ is considerably larger compared to those in $A_2$IrO$_3$ ($A=$ Na and Li) and $\alpha$-RuCl$_3$~\cite{PhysRevLett.113.107201, KI2016, WI2016}, but it is rapidly reduced by $U_p$ and becomes comparable to those for $U_p=1.0$~eV. 
This appears to justify the inclusion of $U_p$ to compensate the less screening in the present MLWF analysis including the O $2p$ orbitals. 
The $d$-$p$ energy differences are estimated as $\Delta_{d-p_{x,y}} \simeq 2.75$~eV and $\Delta_{d-p_z} \simeq 0.92$~eV for the $p_{x,y}$ and $p_z$ orbitals (almost independent of the $t_{2g}$ orbitals), respectively. 
We note that $\Delta_{d-p_z}$ is rather small and in a similar energy scale of the $d$-$p$ transfers, which might hamper the perturbation theory, but the inclusion of $U_p$ reconciles this situation. 
In addition, the small $\Delta_{d-p_z}$ suggests that further-neighbor transfers can be relevant through the indirect transfers. 
Indeed, our MLWF analyses find that the second- and third-neighbor transfer integrals, which arise dominantly from the $d$-$p$-$p$-$d$ indirect transfers, include the matrix elements whose magnitudes are comparable to the nearest-neighbor ones at $U_p=0$. 
Note, however, that the values are more rapidly reduced by $U_p$ than the nearest-neighbor ones, as they are higher-order contributions in the perturbation theory.

We obtain similar results for ZnIrO$_3$. 
The results are summarized in Appendix~\ref{appA}. 
It is noted that the nearest-neighbor $xy$-$yz$ transfer is one order of magnitude larger for ZnIrO$_3$ compared to that for MgIrO$_3$. 
This is presumably due to the larger buckling of the Ir honeycomb planes in ZnIrO$_3$.

\subsubsection{Effective interaction between $j_{\rm eff}=1/2$ pseudospins}
\label{subsubsec:effective interaction} 

Using the perturbation expansion from the atomic limit of the multiorbital Hubbard model based on the MLWF analysis in Table~\ref{tab:tMgIrO3}, we derive an effective model for the $j_{\rm eff}=1/2$ pseudospin degree of freedom (see Sec.~\ref{sec:method}). 
The effective pseudospin Hamiltonian on the nearest-neighbor $z$ bond is summarized as
\begin{equation}
\mathcal{H}^{ \left( z \right) }_{ij}=
\mathbf{S}_i^{\rm T}
\begin{bmatrix} 
J & \Gamma & \Gamma' \\
\Gamma & J & \Gamma' \\
\Gamma'  & \Gamma' & J+K
\end{bmatrix}
\mathbf{S}_{j},
\label{eq:Heff}
\end{equation}
where $\mathbf{S}_i=(S_i^x,S_i^y,S_i^z)^{\rm T}$ denotes the pseudospin operator at site $i$; $J$, $K$, $\Gamma$, and $\Gamma'$ denote the coupling constants for the Heisenberg, the Kitaev, and the two different types of symmetric off-diagonal interactions, respectively.

\begin{figure}[t!]
\includegraphics[width=0.95\columnwidth]{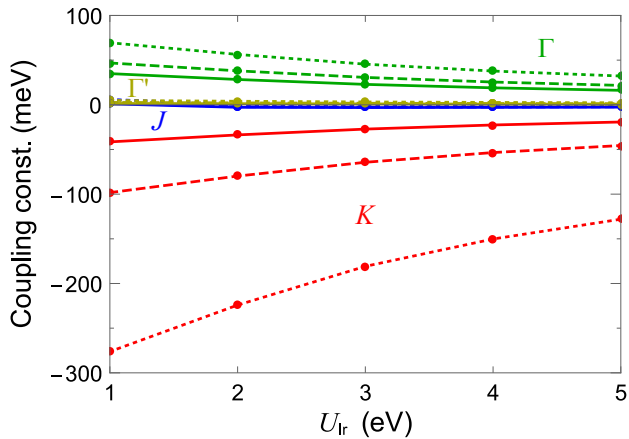} 
\caption{\label{fig:MgIrO3_coupling_const}
Coupling constants for the nearest-neighbor pseudospins in MgIrO$_3$ as functions of the Coulomb repulsion at the Ir site, $U_{\rm Ir}$; see Eq.~\eqref{eq:Heff}. 
The Hund's-rule coupling $J_{\rm H}$ and the spin-orbit coupling coefficient $\lambda$ are set to $J_{\rm H}/U_{\rm Ir}=0.1$ and $\lambda = 0.4$~eV, respectively. 
The data connected by the blue, red, green, and yellow lines represent the Heisenberg $J$, Kitaev $K$, and symmetric off-diagonal couplings $\Gamma$ and $\Gamma'$, respectively; 
the dotted, dashed, and solid lines indicate the data obtained by taking $U_p=0.0$, $0.5$, and $1.0$~eV, respectively.
}
\end{figure} 

The coupling constants estimated for MgIrO$_3$ are plotted in Fig.~\ref{fig:MgIrO3_coupling_const} as functions of $U_{\rm Ir}$ with $J_{\rm H}/U_{\rm Ir}=0.1$ 
and the spin-orbit coupling coefficient $\lambda=0.4$~eV~\cite{PhysRevLett.113.107201}. 
The three lines for each coupling constant display the results for $U_p=0.0$, $0.5$, and $1.0$~eV.
We find that the ferromagnetic Kitaev interaction $K$ is always predominant, and the symmetric off-diagonal interaction $\Gamma$ is subdominant; the Heisenberg interaction $J$ and the other symmetric off-diagonal interaction $\Gamma'$ are vanishingly small. 
This means that the low-energy magnetic property of the spin-orbit coupled Mott insulating state in MgIrO$_3$ is well described by the model with $K$ and $\Gamma$ for the nearest-neighbor sites. 
The model is called the $K$-$\Gamma$ model and has been studied in the context of the Kitaev spin liquid, especially for one of the candidates $\alpha$-RuCl$_3$~\cite{RA2014, RU2019}. 

We note that the magnitude of $K$ is significantly large compared to those for $A_2$IrO$_3$ ($A=$ Na and Li)~\cite{CH2012, PhysRevLett.113.107201, Katukuri_2014, Williams2016, WI2016, Gupta2016} when we assume $U_p=0$, mainly due to the contributions from the large $yz$-$zx$ transfer in Table~\ref{tab:tMgIrO3}. 
However, all the coupling constants are substantially reduced by taking into account $U_p$, as shown in Fig.~\ref{fig:MgIrO3_coupling_const}, according to the reduction of the effective transfer integrals. 
For instance, for $U_p=1$~eV and $U_{\rm Ir}=3$~eV, the value of $K$ is reduced to $-27.1$~meV, which is comparable to that for $A_2$IrO$_3$.
Although the proper values of $U_p$ and $U_{\rm Ir}$ are unknown, the important conclusion is that the nearest-neighbor magnetic interactions in MgIrO$_3$ can be well described by the $K$-$\Gamma$ model irrespective of $U_p$ and $U_{\rm Ir}$.

As discussed in Sec.~\ref{subsubsec:transfer}, there are substantial further-neighbor transfers through the indirect contributions via the O $2p$ orbitals. 
They give rise to sizable further-neighbor exchange interactions, while the coupling constants are reduced by $U_p$ more quickly than the nearest-neighbor ones as they are higher-order processes. 
For instance, assuming $U_{\rm Ir}=3$~eV and $U_p=1$~eV, the dominant second-neighbor contributions within the same honeycomb layer are the antiferromagnetic $K\simeq 10.3$~meV and the ferromagnetic $J\simeq -8.05$~meV, while the dominant third-neighbor one is the antiferromagnetic $J\simeq 9.06$~meV. 
We note that the second-neighbor bonds do not have the inversion centers, and hence, include the subdominant Dzyaloshinskii-Moriya interaction, whose energy scale is estimated as $\simeq 5.76$~meV. 
In addition, we expect contributions from the interlayer couplings through the TiO$_6$ layer.
We speculate that the rather small value of the Curie-Weiss temperature $-67.1$~K in MgIrO$_3$ could be accounted for by a balance among the exchange couplings including such further-neighbor contributions~\cite{PhysRevMaterials.2.054411}.
While the magnetic structure in the ordered phase is experimentally unknown thus far, it will also be determined under the competing exchange interactions; it is left for future study to precisely construct the effective pseudospin model by determining the values of $U_{\rm Ir}$, $J_{\rm H}$, and $U_p$, and to investigate the stable magnetic structure in the ground state.

\begin{figure}[t] 
\includegraphics[width=0.95\columnwidth]{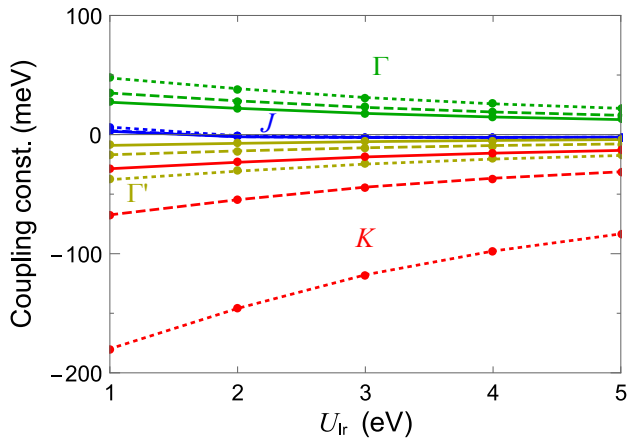} 
\caption{\label{fig:ZnIrO3_coupling_const}
Coupling constants for the nearest-neighbor pseudospins in ZnIrO$_3$ as functions of $U_{\rm Ir}$ with $J_{\rm H}/U_{\rm Ir}=0.1$ and $\lambda=0.4$~eV. 
The notations are common to those in Fig.~\ref{fig:MgIrO3_coupling_const}.
}
\end{figure} 

In the case of ZnIrO$_3$, we plot the effective coupling constants in Fig.~\ref{fig:ZnIrO3_coupling_const}. 
The result indicates that, similar to MgIrO$_3$, the magnetic exchange interactions between the neighboring pseudospins are well described by the dominant $K$ and the subdominant $\Gamma$, while the other symmetric off-diagonal interaction $\Gamma'$ has a small but nonnegligible value, in contrast to the case of MgIrO$_3$. 
This is due to the contribution from the $xy$-$yz$ transfer discussed in Sec.~\ref{subsubsec:transfer}. 

\subsection{MnIrO$_3$}
\label{subsec:MnIrO3}

\begin{table}[t] 
\caption{\label{tab:MnIrO3}
Structural parameters of the optimized lattice structure of MnIrO$_3$ with the trigonal $R\bar{3}$ symmetry: the lattice constants $a$ and $c$ for the conventional unit cell shown in Fig.~\ref{fig:ilmenite}(a), the Wyckoff positions of the Mn, Ir, and O ions, and the bond distances $d$ and angles $\theta$ for neighboring ions within the same honeycomb layer.
}
\begin{ruledtabular}
\begin{tabular}{cc}
$a$ (\si{\angstrom})&$4.9979$\\
$c$ (\si{\angstrom})&$13.159$\\
\hline
Mn ($6c$)&($0$, $0$, $0.34535$)\\
Ir ($6c$)&($0$, $0$, $0.15473$)\\
O ($18f$)&($0.35348$, $0.010833$, $0.077788$)\\
\hline
$d_{\rm Mn-Mn}$ (\si{\angstrom})&$2.9028$\\
$d_{\rm Mn-O}$ (\si{\angstrom})&$1.8828$, $1.9803$\\
$\theta_{\rm Mn-O-Mn}$ (\si{\degree})&$97.398$\\
\hline
$d_{\rm Ir-Ir}$ (\si{\angstrom})&$2.9026$\\
$d_{\rm Ir-O}$ (\si{\angstrom})&$2.0133$, $2.0700$\\
$\theta_{\rm Ir-O-Ir}$ (\si{\degree})&$90.596$\\
\end{tabular}
\end{ruledtabular}
\end{table}

\begin{figure}[t]
\includegraphics[width=1.0\columnwidth]{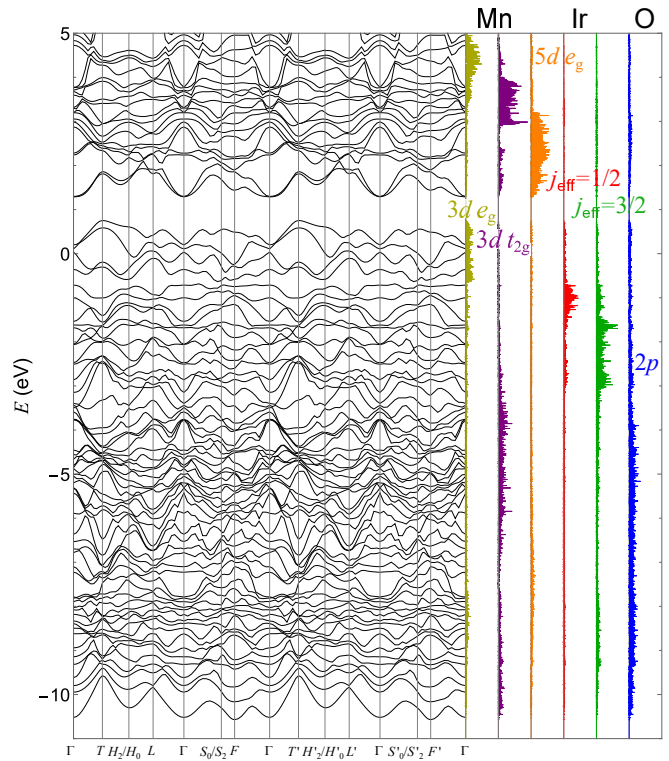} 
\caption{\label{fig:MnIrO3_band_U=3}
The electronic band structure of MnIrO$_3$ obtained by the LDA+SOC+$U$ calculations for the state with $c$-FM for Mn and $ab$-FM for Ir with $U_{\rm Mn}=6$~eV and $U_{\rm Ir}=3$~eV. 
The right panels display the projected DOS for each orbital.
The Fermi level is set to zero.
}
\end{figure} 

Finally, we discuss the fictitious compound MnIrO$_3$. 
Since this compound has not been synthesized thus far, we perform the structural optimization starting from the lattice structure of MnTiO$_3$, as described in Sec.~\ref{sec:method}. 
The optimal structural data are shown in Table~\ref{tab:MnIrO3}. 
We find that the structure of MnIrO$_3$ is significantly distorted from that of MnTiO$_3$.
In particular, the Mn-O bond lengths $d_{\rm Mn-O}$ are largely contracted from $d_{\rm Mn-O} = 2.1060$~\si{\angstrom} and $2.3026$~\si{\angstrom} for MnTiO$_3$ to $1.8828$~\si{\angstrom} and $1.9803$~\si{\angstrom}, and the Mn-O-Mn bond angle $\theta_{\rm Mn-O-Mn}$ is largely widened from $\theta_{\rm Mn-O-Mn} = 87.956$\si{\degree} for MnTiO$_3$ to $97.398$\si{\degree}, as shown in Table~\ref{tab:MnIrO3}. 

We study the electronic band structure for the optimal lattice structure by the LDA+SOC+$U$ calculation.  
We assume $U_{\rm Mn}=6$~eV and $U_{\rm Ir}=3$~eV. 
From energy comparison between different magnetic states for Mn and Ir sites, we find that the state with $c$-FM for Mn and $ab$-FM for Ir has the lowest energy, while the magnetic moments are reduced from the MnTiO$_3$ and MgIrO$_3$ cases: 
$|\mathbf{m}_{\rm Mn}| \simeq 3.4$~$\mu_{\rm B}$ and $|\mathbf{m}_{\rm Ir}| \simeq 0.008$~$\mu_{\rm B}$.
The second lowest energy is the state with $c$-AFM for Mn ($|\mathbf{m}_{\rm Mn}| \simeq 3.4$~$\mu_{\rm B}$) and $ab$-AFM for Ir ($|\mathbf{m}_{\rm Ir}| \simeq 0.005$~$\mu_{\rm B}$), whose energy is higher by ${\sim} 69$~meV. 

The electronic structure for MnIrO$_3$ with $c$-FM for Mn and $ab$-FM for Ir is shown in Fig.~\ref{fig:MnIrO3_band_U=3}. 
We find that, in contrast to MnTiO$_3$, MgIrO$_3$, and ZnIrO$_3$, the system is metallic and the relevant bands near the Fermi level are composed of the hybridized ones between the Mn $3d$, Ir $5d$, and O $2p$ orbitals.
Notably, we find that the valence of Ir ions is considerably different from those for MgIrO$_3$ and ZnIrO$_3$. 
By integrating the projected DOS below the Fermi level, we obtain Ir$^{2.02+}$ ($6.98$ $5d$ electrons per Ir ion) for MnIrO$_3$, which is far from Ir$^{3.66+}$ ($5.34$ $5d$ electrons) for MgIrO$_3$ and ZnIrO$_3$ at $U_{\rm Ir} = 3$~eV.
These results suggest that the low-energy physics of the Ir honeycomb network in MnIrO$_3$ is not properly described by the effective $j_{\rm eff}=1/2$ pseudospins which are expected for Ir$^{4+}$.
On the other hand, we find that the valence of Mn ions in MnIrO$_3$ is similar to that for MnTiO$_3$: Mn$^{2.05+}$ for MnIrO$_3$ and Mn$^{1.89+}$ for MnTiO$_3$, both of which are close to Mn$^{2+}$.

\section{Summary}
\label{sec:summary}

To summarize, we have studied the electronic and magnetic properties of the iridium ilmenites MgIrO$_3$ and ZnIrO$_3$, in comparison with the conventional antiferromagnetic insulator MnTiO$_3$. 
From the {\it ab initio} calculations, we showed that both Ir compounds have typical electronic band structures of the spin-orbital coupled Mott insulator: the low-energy Ir $5d$ bands are split into the $j_{\rm eff}=1/2$ doublet and the $j_{\rm eff}=3/2$ quartet by the spin-orbit coupling, and the half-filled $j_{\rm eff}=1/2$ doublet is further split by the Coulomb interaction. 
By using the multiorbital Hubbard model obtained by the MLWF analysis of the band structure, we found that the low-energy magnetic properties are well described by the $j_{\rm eff}=1/2$ pseudospins interacting with the predominant Kitaev-type bond-dependent interaction and the subdominant symmetric off-diagonal interactions;
more specifically, MgIrO$_3$ and ZnIrO$_3$ are well approximated by the $K$-$\Gamma$ and $K$-$\Gamma$-$\Gamma'$ models, respectively, while further-neighbor contributions are expected to be relevant as well. 
In addition, we calculated the electronic band structure for the fictitious compound MnIrO$_3$ with structural optimization, and showed that it does not provide the $j_{\rm eff}=1/2$ physics because of the metallic nature and the different ionic state of Ir. 

Our results indicate that the iridium ilmenites MgIrO$_3$ and ZnIrO$_3$ offer a good platform for exotic magnetism described by the spin-orbital entangled $j_{\rm eff}=1/2$ moments. 
The importance of $\Gamma$ as well as $\Gamma'$ suggests a similarity to $\alpha$-RuCl$_3$ rather than $A_2$IrO$_3$ ($A=$ Na and Li), probably due to structural similarity in the lack of cations at the hexagon centers in the honeycomb layers. 
However, the magnitudes of the coupling constants would be much larger than those for the $4d$-electron compound $\alpha$-RuCl$_3$ and comparable to those for the $5d$-electron compounds $A_2$IrO$_3$, due to the larger spatial extent of the electron wave functions and the weaker Coulomb interactions. 
Thus, the Ir ilmenites are the materials that inherit the structural aspect from $\alpha$-RuCl$_3$ and the electronic aspect from $A_2$IrO$_3$. 
Our findings would not only be helpful to understand the magnetism in these compounds but also provide a guide toward the realization of the Kitaev spin liquid by designing the magnetic interactions. 

\begin{acknowledgments}
The crystal structures in Figs.~\ref{fig:ilmenite}(a) and \ref{fig:ilmenite}(b) were visualized by \texttt{VESTA}~\cite{MO2011}. 
Ref.~\onlinecite{HINUMA2017140} was referred to for the Brillouin zone and the symmetric points in Fig.~\ref{fig:ilmenite}(c).
Parts of the numerical calculations have been done using the facilities of the Supercomputer Center, the Institute for Solid State Physics, the University of Tokyo. 
This work was supported by JST CREST (JP-MJCR18T2), and JSPS KAKENHI Grants No.~19H05825 and No.~20H00122. 
\end{acknowledgments}

\appendix

\section{ZnIrO$_3$}
\label{appA}

\begin{figure}[ht!]
\includegraphics[width=0.85\columnwidth]{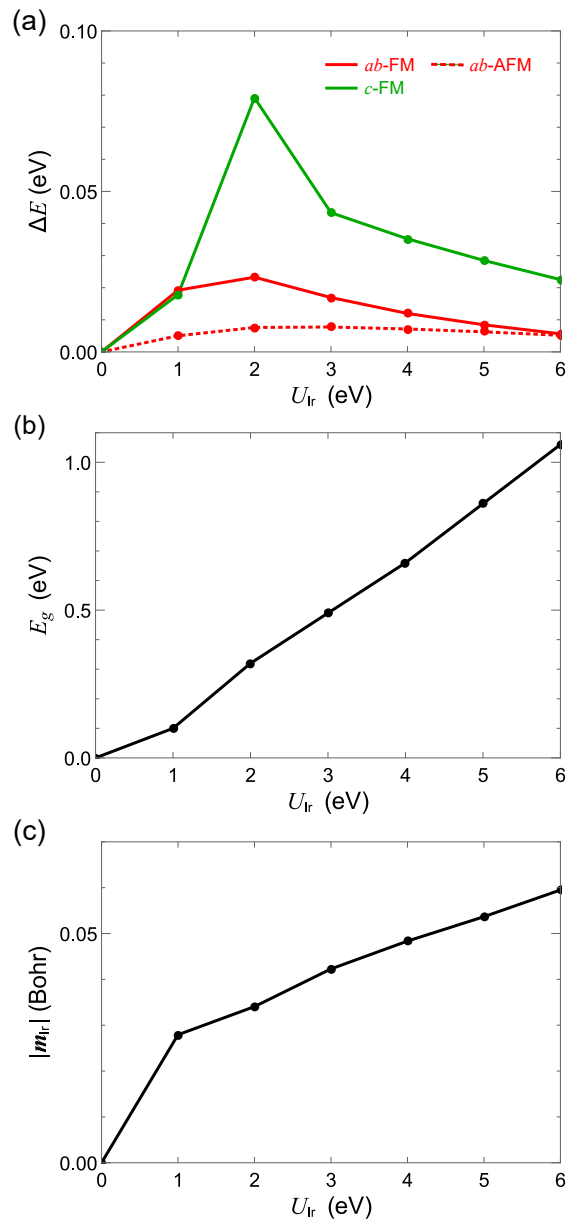}
\caption{\label{fig:ZnIrO3_Udep}
(a) Energy measured from the $c$-AFM state in ZnIrO$_3$ as a function of the Coulomb repulsion at the Ir site, $U_{\rm Ir}$, obtained by the LDA+SOC+$U$ calculations.
(b) The energy gap and (c) the antiferromagnetic moment at the Ir site in the $c$-AFM state.
}
\end{figure} 

\begin{figure}[t]
\includegraphics[width=1.0\columnwidth]{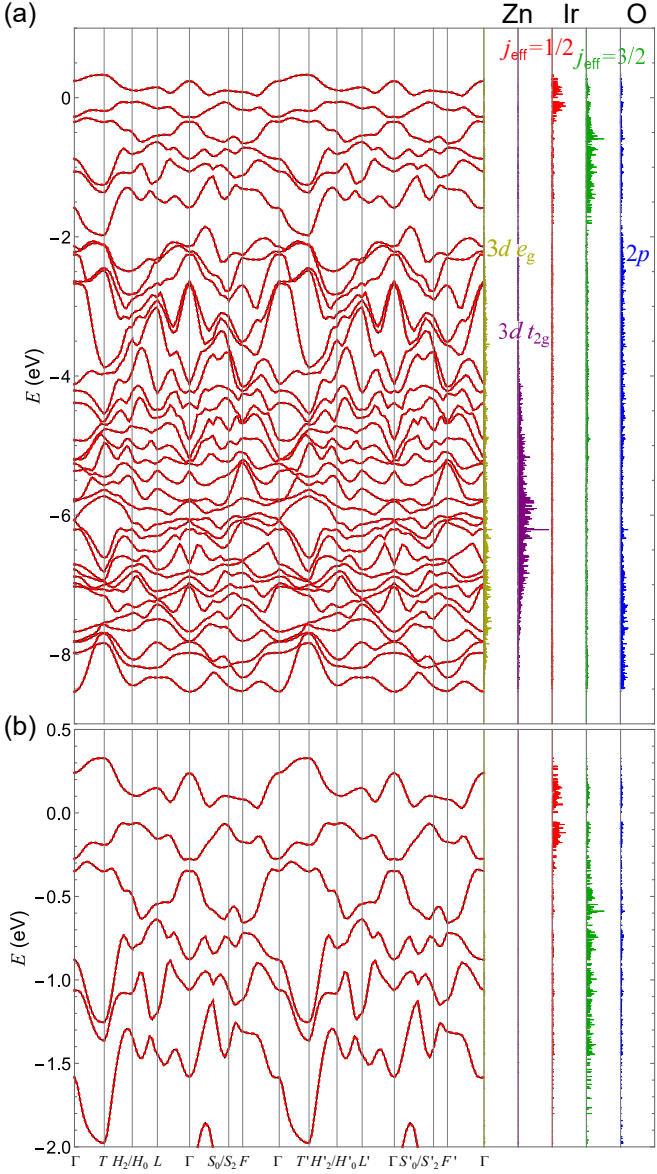}
\caption{\label{fig:ZnIrO3_band_U=0}
The electronic band structure of ZnIrO$_3$ obtained by the LDA calculations ($U_{\rm Ir}=0$) for the paramagnetic metallic state. 
The notations are common to those in Fig.~\ref{fig:MgIrO3_band_U=0}.
}
\end{figure} 

In this Appendix, we present the results for ZnIrO$_3$. 
Figure~\ref{fig:ZnIrO3_Udep}(a) shows the energy comparison between different magnetic states. 
Similar to MgIrO$_3$ in Sec.~\ref{subsec:MgTiO3 and ZnIrO3}, ZnIrO$_3$ is a paramagnetic metal at $U_{\rm Ir}=0$, whose electronic band structure is similar to that for MgIrO$_3$ as shown in Fig.~\ref{fig:ZnIrO3_band_U=0}. 
Upon introducing $U_{\rm Ir}$, however, the lowest-energy state is the $c$-AFM state in contrast to the $ab$-AFM in MgIrO$_3$, while the energy difference $\Delta E$ to the second-lowest state, the $ab$-AFM, is not so large (less than $0.01$~eV), as shown in Fig.~\ref{fig:ZnIrO3_Udep}(a). 
The result indicates that ZnIrO$_3$ has an out-of-plane magnetic anisotropy, although the experiment shows an easy-plane anisotropy~\cite{PhysRevMaterials.2.054411}. 
Nonetheless, the behaviors of the energy gap and the antiferromagnetic moment are similar to those for MgIrO$_3$, as shown in Figs.~\ref{fig:ZnIrO3_Udep}(b) and \ref{fig:ZnIrO3_Udep}(c), respectively. 

\begin{figure}[t]
\includegraphics[width=1.0\columnwidth]{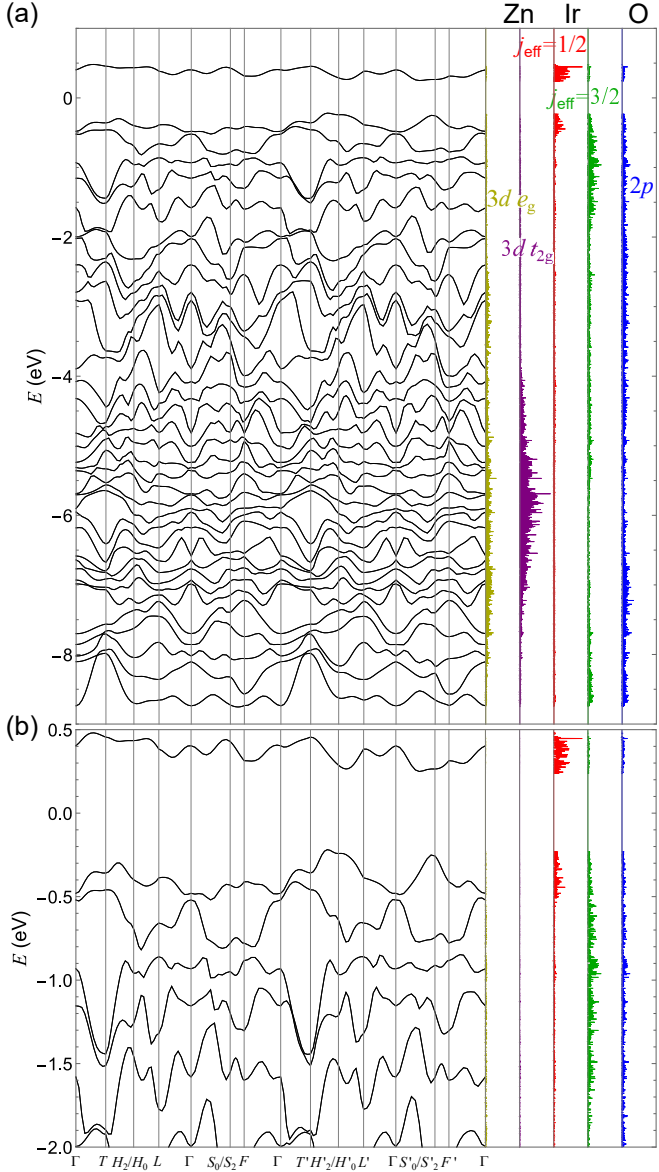} 
\caption{\label{fig:ZnIrO3_band_U=3}
The electronic band structure of ZnIrO$_3$ obtained by the LDA+SOC+$U$ calculations for the $c$-AFM state with $U_{\rm Ir}=3$~eV.
The notations are common to those in Fig.~\ref{fig:MgIrO3_band_U=3}.
}
\end{figure} 

\begin{widetext}
\begin{table*}[th!]
\caption{\label{tab:tZnIrO3}
Transfer integrals between the Ir $t_{2g}$ orbitals on the nearest-neighbor $z$ bond for ZnIrO$_3$. 
The notations are common to those in Table~\ref{tab:tMgIrO3}.
}
\begin{ruledtabular}
\begin{tabular}{ccccccc}
&$yz\uparrow$&$yz\downarrow$&$zx\uparrow$&$zx\uparrow$&$xy\uparrow$&$xy\uparrow$\\
\hline
$yz\uparrow$&\makecell{41.9\\77.0\\94.7}\\
\hline
$yz\downarrow$&\makecell{0.00\\0.00\\0.00}&\makecell{41.9\\77.0\\94.7}\\
\hline
$zx\uparrow$&\makecell{1190-12.0{\rm i}\\765-11.2{\rm i}\\553-10.6{\rm i}}&\makecell{-41.2{\rm i}\\-29.9{\rm i}\\-23.8{\rm i}}&\makecell{41.9\\77.0\\94.7}\\
\hline
$zx\downarrow$&\makecell{-41.2{\rm i}\\-29.9{\rm i}\\-23.8{\rm i}}&\makecell{1190+12.0{\rm i}\\765+11.2{\rm i}\\553+10.6{\rm i}}&\makecell{0.00\\0.00\\0.00}&\makecell{41.9\\77.0\\94.7}\\
\hline
$xy\uparrow$&\makecell{-258+8.05{\rm i}\\-198+5.29{\rm i}\\-166+3.84{\rm i}}&\makecell{-72.6{\rm i}\\-58.6{\rm i}\\-49.9{\rm i}}&\makecell{-258+8.05{\rm i}\\-198+5.29{\rm i}\\-166+3.84{\rm i}}&\makecell{72.6{\rm i}\\58.6{\rm i}\\49.9{\rm i}}&\makecell{-312\\-344\\-366}\\
\hline
$xy\downarrow$&\makecell{-72.6{\rm i}\\-58.6{\rm i}\\-49.9{\rm i}}&\makecell{-258-8.05{\rm i}\\-198-5.29{\rm i}\\-166-3.84{\rm i}}&\makecell{72.6{\rm i}\\58.6{\rm i}\\49.9{\rm i}}&\makecell{-258-8.05{\rm i}\\-198-5.29{\rm i}\\-166-3.84{\rm i}}&\makecell{0.00\\0.00\\0.00}&\makecell{-312\\-344\\-366}\\
\end{tabular}
\end{ruledtabular}
\end{table*}
\end{widetext}

Figure~\ref{fig:ZnIrO3_band_U=3} shows the electronic band structure of the $c$-AFM insulating state for ZnIrO$_3$. 
Again, the result is similar to that for MgIrO$_3$: the energy gap opens between the two $j_{\rm eff}=1/2$ bands, realizing the spin-orbit coupled Mott insulating state. 
In this case, we perform the MLWF analysis including the Zn $3d$ orbitals as mentioned in Sec.~\ref{sec:method}, since the energy levels overlap with those for the O $2p$ orbitals. 
We plot the obtained tight-binding band structure in Fig.~\ref{fig:ZnIrO3_band_U=0}, which well reproduce the {\it ab initio} result. 
In Table~\ref{tab:tZnIrO3}, we show the effective transfer integrals for the Ir $t_{2g}$ orbitals on the nearest-neighbor $z$ bond estimated from the MLWF analysis. 
In the calculations, we use $\Delta_{d-p_{x,y}}\simeq 3.09$~eV and $\Delta_{d-p_z}\simeq 1.02$~eV estimated from the MLWF analysis.
Similar to the case of MgIrO$_3$, the dominant effective transfer is between the $yz$ and $zx$ orbitals. 
Meanwhile, the $xy$-$yz$ transfer has much larger value than that for MgIrO$_3$, presumably due to the larger buckling of the Ir honeycomb planes, as mentioned in the main text; the height difference in the $c$ direction between two neighboring Ir sites is $1.3341 \times 10^{-2} c$ for ZnIrO$_3$, while it is $1.5734 \times 10^{-3} c$ for MgIrO$_3$.
The estimates of the effective coupling constants between the pseudospins are shown in Fig.~\ref{fig:ZnIrO3_coupling_const}.


\bibliography{main}

\end{document}